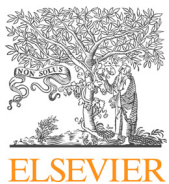
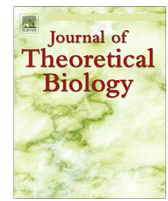

# Persistent homology and the shape of evolutionary games

Jakob Stenseke

*Department of Philosophy, Lund University, Helgonavagen 3, Lund 221 00, Sweden*

ABSTRACT

For nearly three decades, spatial games have produced a wealth of insights to the study of behavior and its relation to population structure. However, as different rules and factors are added or altered, the dynamics of spatial models often become increasingly complicated to interpret. To tackle this problem, we introduce persistent homology as a rigorous framework that can be used to both define and compute higher-order features of data in a manner which is invariant to parameter choices, robust to noise, and independent of human observation. Our work demonstrates its relevance for spatial games by showing how topological features of simulation data that persist over different spatial scales reflect the stability of strategies in 2D lattice games. To do so, we analyze the persistent homology of scenarios from two games: a Prisoner's Dilemma and a SIRS epidemic model. The experimental results show how the method accurately detects features that correspond to real aspects of the game dynamics. Unlike other tools that study dynamics of spatial systems, persistent homology can tell us something meaningful about population structure while remaining neutral about the underlying structure itself. Regardless of game complexity, since strategies either succeed or fail to conform to shapes of a certain topology there is much potential for the method to provide novel insights for a wide variety of spatially extended systems in biology, social science, and physics.

## 1. Introduction

Game theory and its evolutionary extensions have provided extraordinarily productive paradigms for understanding behavior. Nearly seven decades after John Nash extended von Neumanns study of two-player zero-sum games with the introduction of Nash Equilibrium (NE), game theoretical models have become a standard modus operandi in economics, social science, and biology (Nash, 1950; Holt and Roth, 2004). Evolutionary game theory (EGT) broadened the study even further with the addition of evolutionary stable strategies (ESS) and replicator dynamics (Smith and Price, 1973; Taylor and Jonker, 1978). This has in turn allowed us to describe how strategies change over time, and contrary to classical game theory, EGT does not assume perfect rationality (Smith and Price, 1973; Friedman, 1991; van Damme, 1994; Weibull, 1997).

Further on, the intuition that spatial effects have a major impact on evolutionary dynamics has led to the exploration of spatial games (Levin and Paine, 1974; Hassell et al., 1991; Nowak and May, 1992; Nowak and May, 1993; Lindgren and Nordahl, 1994; Szabó and Tke, 1998; Hauert, 2006; Perc and Szolnoki, 2008). These include deterministic and stochastic cellular automata (Nowak and May, 1992; Nowak and May, 1993; Hiebeler, 1997), exceedingly complex spatial patterns (Hassell et al., 1991), open-ended space–time chaos (Lindgren and Nordahl, 1994), and reaction–diffusion models with Turing patterns (Cressman and Vickers, 1997; Wakano et al., 2009). It has been shown that space can influence game dynamics in a variety of ways, e.g., depending on the underlying structure (Lieberman et al., 2005), whether interaction is stochastic or deterministic (Nowak et al., 1994), and whether space is discrete or continuous (Durrett and Levin, 1994). Besides spatial effects, additional factors may be added or modified to influence the evolutionary dynamics; by varying the numbers of possible strategies (Mathiesen et al., 2011), pay-off values (Lindgren and Nordahl, 1994; Perc et al., 2017), update rules (Moyano and Sánchez, 2009), memory strategies (Axelrod and Axelrod, 1984; Lindgren and Nordahl, 1994), social states (Hauert, 2006; Perc and Szolnoki, 2008), and coevolution (Perc and Szolnoki, 2009). While the multitude of approaches to spatial games offer flexible ways to model intricate social and biological behavior, they also pave the way for experimental design bias (Alexander, 2019). To that end, there is a risk that the systems merely reflect our imaginative ways of modeling as opposed to real-world phenomena. Furthermore, in increasingly complex games, the evolutionary dynamics occasionally becomes indistinguishable from random fluctuations, which hinders reproducibility and predictability (Hassell et al., 1991).

*E-mail address:* jakob.stenseke@fil.lu.se

Spatial evolutionary games seem to be caught between trivial simplicity and intractable complexity, ranging from the simple which can be illuminated by a single NE (e.g., the standard Prisoners Dilemma), to more or less unstable situations of multiple NE and ESS, infinite number of NE, or none at all (Samuelson, 1997; Skyrms, 2014). Traditionally, the dynamics of evolutionary games have been specified by the replicator equation, where strategies reproduce proportionally to their payoffs (Taylor and Jonker, 1978). But as the equation assumes uniform distribution of strategies, it fails to account for population structure. Instead, several alternative methods have been proposed to understand evolutionary dynamics and pattern formation in spatial models, most notably mean-field models (Parisi, 1988; Caswell and Cohen, 1991; Hanski, 1991), local structure theory (Gutowitz et al., 1987; Hiebeler, 1997), statistical physics (Perc et al., 2017; Li et al., 2013; Skyrms, 2014), reaction–diffusion models (Wakano et al., 2009; Durrett and Levin, 1994; DeForest and Belmonte, 2013), and evolutionary graph theory (Lieberman et al., 2005). Although these approaches have provided invaluable tools to the study of spatial games, they all lack to some degree in respect to what they explain, and what they assume about spatial correlation, underlying mechanics, and population structure. Tools from statistical physics have illuminated complicated evolutionary dynamics through mean-field approximations, phase diagrams, and Monte Carlo methods, but fail to account for intricate spatial information as such (Perc et al., 2017; Skyrms, 2014; Li et al., 2013). Local structure theory can address local correlations, but only through mean-field approximations of predefined groups of set sizes. Evolutionary graph theory can describe how population structure as such affect evolutionary dynamics (by generalizing population structure), but as it starts from the assumption of some underlying topology, it cannot explain the effects in situations where these structures are unknown. Similarly, while reaction–diffusion models can successfully identify the complex a priori mechanisms that give rise to patterns and spatial heterogeneity a posteriori, they lack a rigorous a priori characterization of the patterns themselves.

To mitigate these problems, we introduce persistent homology (PH) as a novel method to analyze the spatial structure of games in a manner that is robust to noise and insensitive to the choice of parameter values and underlying game mechanics. PH and other techniques from topological data analysis (TDA) have already produced a wealth of insights in an exceedingly diverse set of applications, including cell development (Rizvi et al., 2017), proteins (Kovacev-Nikolic et al., 2016; Gameiro et al., 2015; Xia and Wei, 2014), collective biological behavior (Topaz et al., 2015), DNA structure (Emmett et al., 2016), cancer (Nicolau et al., 2011; DeWoskin et al., 2010; Crawford et al., 2020), molecular stability (Xia et al., 2015), plant morphology (Li et al., 2017), and phylogenetics (Chan et al., 2013; Cámara et al., 2016). The basic motivation is that shape matters, and features that persist over a wide range of spatial scales are assumed to represent real higher-order features of the underlying space (Carlsson, 2009). In relation to evolutionary games, the major benefit is that persisting features are more likely to reflect the stability of strategies rather than being the result of noise, sampling, or specific parameter choices. No matter how simple or complex a game is, strategies either succeed or fail to conform to a particular shape that has a certain topological quality. TDA and PH thus provides a promising "quality from quantity" approach to the study of evolutionary dynamics as it, unlike statistical methods, tells us something meaningful about the population structure while, unlike evolutionary graph theory, reaction–diffusion models, and local structure theory, does not assume anything particular about the underlying mechanism or structure of the system itself.

The paper is structured as follows. First, we give an introduction to topological data analysis and explain how persistent homology is computed and visually represented in barcodes. We then explain an application to spatial games by describing three general topological features that correspond to the stability of strategies in a 2D lattice game. Specifically, by stacking iterations of 2D states, we demonstrate how the spatio-temporal information of a game can be analyzed as a topological object. Then we create two evolutionary games: a Prisoners Dilemma game with stochastic interaction and reproduction rules, and a cyclical SIRS disease model. We then analyze the persistent homology of a number of scenarios of the games and present the results. Finally, we discuss the major benefits and challenges of the approach and identify some promising venues for future work. In particular, by combining and comparing PH with other methods, we describe the theoretical significance and practical utility it can provide to the study of pattern formation.

## 2. Method

### 2.1. TDA and persistent homology

TDA can either refer to a broad class of data analysis methods that make use of topological notions such as connectivity and shape (see Wasserman, 2018 for a recent review), or more narrowly to the particular method of persistent homology (Zomorodian and Carlsson, 2005; Edelsbrunner and Morozov, 2014). The main aim of TDA is to develop tools to study qualitative features of data, and in order to achieve this, we need clear definitions of what qualitative features are, techniques to compute them, and some assurance regarding their robustness. It is no surprise that the most prominent tool of TDA is persistent homology, as it accommodates all three matters; it is based on rigorous concepts from algebraic topology, is robust to noisy data, and can be computed through linear algebra (Otter et al., 2017).

Simply put, PH extends the link between a topological space and its homology groups to that of the relationship between a function and its persistence diagram. Here, homology refers to a way of associating algebraic objects to other mathematical objects in terms of connectivity, and should not be confused with biological homology, that is, similarity of features based on shared ancestry. The underlying motivation for homology is that shapes can be differentiated by inspecting their holes; a disk and a circle are not homologous as the disk is solid and the circle has a hole. Qualitative features are similarly defined as topological invariants, i.e., features that persist even if the underlying space is deformed by stretching or bending. PH centers around one of the most basic topological invariants, namely the Betti number. For a reasonable topological space, $S$, we can compute a sequence of integers $\{b_m\}_{m \geq 0}$ such that $b_m$ counts the number of $m$-dimensional holes of $S$. Dimension 0 specifies the number of connected components in $S$, 1 the number of holes, 2 the number of voids, and 3 and beyond gives higher-dimensional variations of holes. For instance, a circle and a square have $b_1 = 1$ if they both have an individual 1-dimensional hole, whereas a figure-eight has $b_1 = 2$ since it has two distinct holes. A hollow sphere, on the other hand, has a 2-dimensional void inside it ($b_2 = 1$) but no 1-dimensional holes ($b_1 = 0$). Fig. 1 shows the Betti number of six basic objects.

In order to identify the Betti numbers of a collection of data points $X$ in some finite metric space, we first need to convert it into a global object. The most straight-forward way to do so is to connect the points by proximity, specified by a distance $\epsilon$. More specifically, we construct simplicial complexes; combinatorial structures counting the Betti numbers at one particular distance $\epsilon$ is not sufficient. If $\epsilon$ is too small, the complex is simply a discrete set without connections, and we cannot distinguish real features from noise. On the other hand, if $\epsilon$ is too large, any two points connect, and we end up with one giant high-dimensional simplex. The notion of persistence offers an appropriate response to this problem (Edelsbrunner et al., 2002; Zomorodian and Carlsson, 2005). Instead of choosing a specific $\epsilon$, we consider a wide range of different distances and record the lifetime of features as they appear and disappear at different $\epsilon$. Short-lived features represent noise, while features that persist over a significant range of parameters are assumed to represent real higher-order features of the underlying space. Intuitively, a simplicial complex constructed from the data given some small $\epsilon$ is a subset of the simplicial complex constructed for a larger $\epsilon$ of the same data. More technically, if we consider $(\mathcal{VR}_i)_1^n$ as a sequence of Vietoris-Rips complexes fixed to a set of data points $X$ with a sequence of increasing values $(\epsilon_i)_1^n$, there are natural inclusion maps:

$$\mathcal{VR}_1 \stackrel{\mathcal{M}}{\hookrightarrow} \mathcal{VR}_2 \stackrel{\mathcal{M}}{\hookrightarrow} \cdots \stackrel{\mathcal{M}}{\hookrightarrow} \mathcal{VR}_{n-1} \stackrel{\mathcal{M}}{\hookrightarrow} \mathcal{VR}_n$$

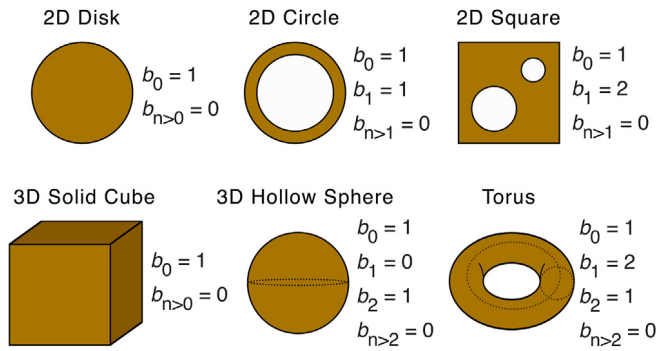

**Fig. 1.** Six basic objects and their Betti-numbers. Simply put, Betti number $m$ counts the $m$-dimensional holes in the topological space.

If we examine the homology ($\mathcal{H}_*$) of these iterated inclusions $\mathcal{M}: \mathcal{H}_*\mathcal{VR}_i \to \mathcal{H}_*\mathcal{VR}_j$, the maps unveil the persisting features of $X$ for all parameter values $i < j$.

In the TDA literature, simplicial complexes and their inclusion maps are usually called filtration, and applying homology to a filtration, we acquire an algebraic structure called a persistence module. Additionally, The Structure Theorem for PIDs implies that a persistence module $PM$ can be decomposed into a direct sum of simple modules, offering a birth and death-pairing of features in $PM$. This has in turn motivated the construction of persistence barcodes, where horizontal lines represent the appearance and disappearance of features in different dimensions as one increase the distance $\epsilon$. A persistence barcodes is essentially a graphical representation of an algebraic structure, offering a compact and effective summary of the variations of invariants over different scales.

To illustrate, we take the set of eleven points as shown in Fig. 2. For different values of $\epsilon$, we create a filtered VR complex; a chain of nested spaces made out of lines, vertices, and triangles (top of the figure). The eleven red lines at the bottom represent the lifetime of the individual components of Betti dimension 0, and the blue dashed lines represent holes of dimension 1. Increasing the distance $\epsilon$, we notice that some components disappear as they connect with others. Around $\epsilon = 0.8$, every point has merged into a single component. At the same distance, we see the first appearance of a hole that persists all the way to $\epsilon = 1.4$. This means that made of simplices. A 0-simplex is a point, a 1-simplex is a line connecting two 0-simplices, and a 2-simplex is a triangle connecting three 0-simplices at each vertex and two 1-simplices at each edge.

The two most used methods to convert data points into global topological objects are the Čech and Vietoris-Rips complexes. Given a set of points $X$, we construct the Čech complex, $\mathcal{C}_\epsilon$, by drawing $\epsilon/2$-balls around each subset $S \subset X$, and include $S$ as a simplex if there is a common point of intersection in the $\epsilon/2$-ball neighborhood of $S$. If $X$ is in a euclidean space, the Nerve Theorem ensures that $\mathcal{C}_\epsilon$ has the same homology as the union of $X$ enclosed by the $\epsilon/2$-ball radius (Edelsbrunner and Harer, 2010). Essentially, this entails that the abstract simplicial complex $\mathcal{C}_\epsilon$ inherits the underlying structure of $X$.

However, computing Čech complexes is costly given the large number of intersections one has to check; in order to tell if there are any $k$-simplices, one has to inspect all subsets of size $k$. It is therefore more suitable to use the less expensive Vietoris-Rips complex, which preserves the relevant topological relationship to the data set by approximating the Čech complex. Given a set of points $X$, we construct the Vietoris-Rips complex, $\mathcal{VR}_\epsilon$, by only including the pairwise intersections of points within $\epsilon$. Since the VR complex includes an edge for every pair, the combinatorics of its 1-skeleton exhaustively determines the complex. Consequently, the VR complex can conveniently be represented as a graph, as opposed to the Čech complex which require us to store the entire boundary operator (Ghrist, 2008).

Having converted a collection of data points to a global topological object, we now turn to the challenge of identifying its relevant topological features, i.e., its Betti numbers. We note that simply the data set has a long-living hole, which we intuitively can confirm by simply observing the spatial distribution of the points. The shorter dashed line, on the other hand, can be interpreted as noise.

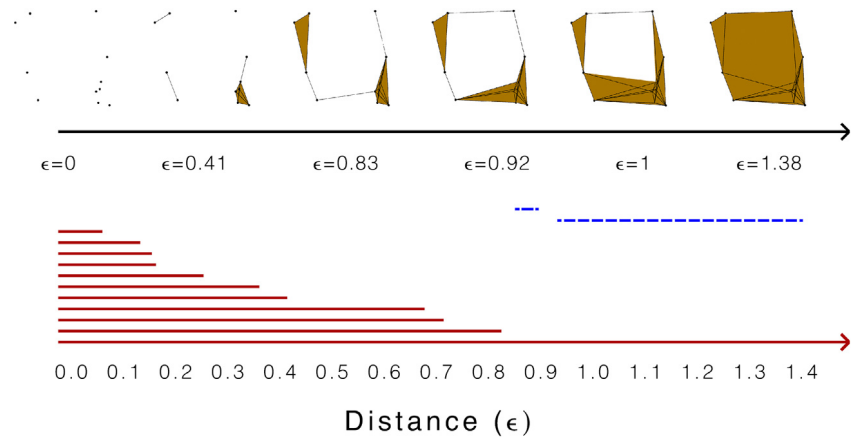

**Fig. 2.** Example of persistent homology applied to a set of data points. The top shows the nested sequence of spaces derived from the data as we increase $\epsilon$ from 0 to 1.4. The bottom shows the persistence barcode of the data structure, representing the lifetime of individual components (red lines) and holes (dashed blue lines) over different $\epsilon$.

The application of PH has provided a wealth of interesting results in a variety of biological fields. By creating connections between data points across different scales, it allows us to both characterize and automatically detect features of any $n$-dimensional space, be it an individual cell (Rizvi et al., 2017), a plant (Li et al., 2017), or a phylogenetic tree (Chan et al., 2013). It is particularly useful to apply PH to data where connections, holes, and voids play an important role for understanding the studied phenomenon. To that end, PH has been used to detect group alignment and clustering events in biological aggregation models (Topaz et al., 2015), quantify the variability of zebrafish patterns in silico (McGuirl et al., 2020), automatically characterize and compare aspects of collective cell motion (Bonilla et al., 2020), model and quantify protein flexibility (Xia and Wei, 2014), describe atomic configurations in amorphous solids (Hiraoka et al., 2016), and characterize the both clonal and reticulate evolution of RNA viruses (Chan et al., 2013).

The computation of PH is under active development, with increasingly faster algorithms that use many innovative shortcuts and optimizations. In the rest of the paper, we will use the TDA-Stats pipeline to compute Vietoris-Rips persistence barcodes (Wadhwa et al., 2018). It uses the powerful Ripser algorithm (Bauer, 2019), which in a recent benchmark outperforms other algorithms by many orders of magnitude (Otter et al., 2017). For a more in-depth introduction to PH, barcodes, and the computation of PH, see (Zomorodian and Carlsson, 2005; Ghrist, 2008; Edelsbrunner and Morozov, 2014; Otter et al., 2017).

*2.2. Persistent homology applied to evolutionary games*

We now describe how persistent homology can be used to analyze the data of evolutionary games. The basic assumption is that spatial games have a shape that reflects the status of the strategies. To take a trivial example, let us consider a single iteration of a 2D game on a lattice with two strategies, $A$ and $B$, as shown in Fig. 3(a). Every space of the 10x10 lattice is occupied by a strategy, from the first bottom-left $P_1(0,0)$, to the last top-right $P_{100}(9,9)$. Intuitively the snapshot tells us that three patches of $B$ (red) exist in a lattice otherwise occupied by $A$ (blue).

Essentially, what PH offers is a way to automate this very intuition (Fig. 3(b)). Using the Cartesian coordinates $(x,y)$ of every $A$ as data, it tells us that three distinct holes persist between distance $\epsilon = 1.41$, corresponding to the euclidean distance between point $P_1(0,0)$ and point $P_2(1,1)$, and $\epsilon = 2$. Note that 0-dimensional features below $\epsilon = 1$ only shows that there are 100 individual components, which is due to the fact that 1 is the minimum distance between any two points. Similarly, 1-dimensional features between $\epsilon = 1$ and 1.41 merely reflects the interim holes that appear when the diagonal distance is not considered.

Although analyzing a single iteration of a game can be fruitful in certain cases, a more interesting application of PH is to add another dimension, namely time, as it allows us to examine the evolutionary dynamics of the game as the strategies change over the course of the game. To do so, we take the 3-dimensional coordinates $(x,y,z)$, where $z$ represents time, of a certain strategy $S$ at every iteration of an evolutionary game as data points. For instance, a data point $S(5,5,5)$ means that strategy $S$ exists at the coordinate $(5,5)$ at iteration 5. Another practical alternative that reduces the computational cost is to use the 3-dimensional coordinates at every iteration interval of particular interest, for instance, by excluding iterations where the game is in a frozen state.

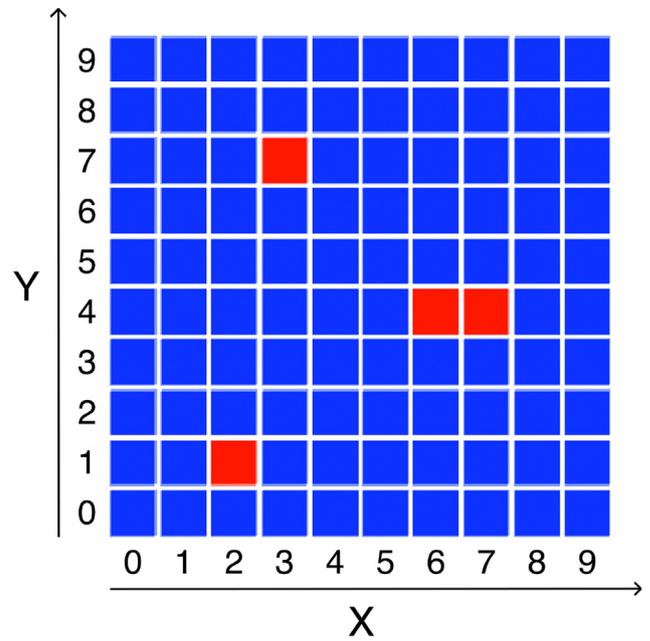

(a)

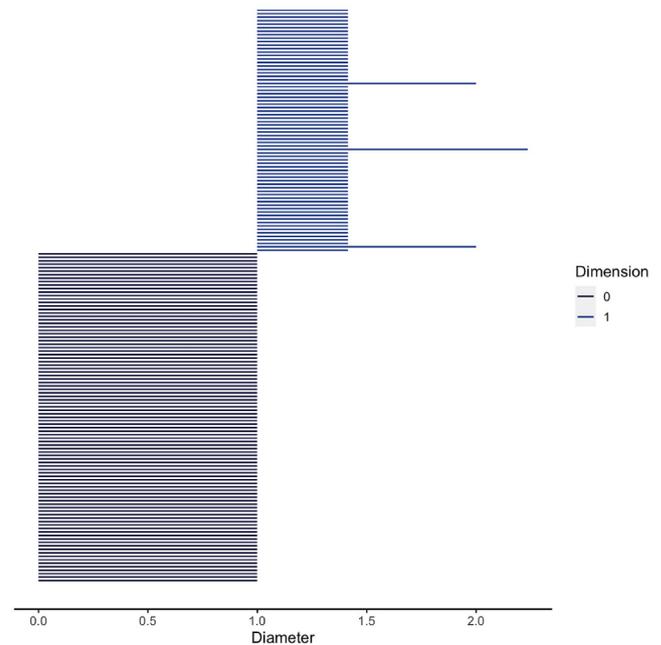

(b)

**Fig. 3.** A single iteration of a spatial game (a) and the persistent homology analysis of it (b). Taking the coordinates $(x,y)$ of every player of strategy $A$ (blue) as data set, it reveals three significant holes persisting from $\epsilon = 1.41$ to $\epsilon = 2$.

By including time in our data, we can now specify three shapes of particular interest that corresponds to the stability of strategies, as shown in Fig. 4.

**Shape (1)** A solid structure where $b_0 = 1$ and $b_{n>0} = 0$ is an evolutionary stable strategy without any other strategies within it (Fig. 4(1)). Simply put, a space that is occupied only by a single strategy over the entire course of the game, or during relevant iteration intervals, is an evolutionary stable strategy as it reflects the simple fact that no other strategy has successfully invaded the space.

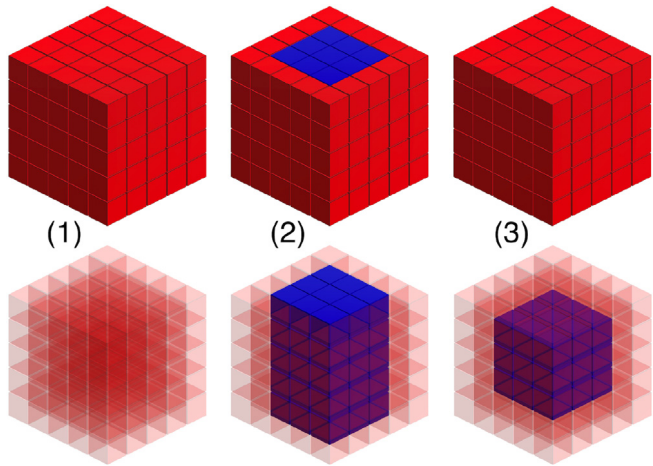

**Fig. 4.** Three basic shapes of topological interest: (1) a single evolutionary stable strategy, (2) a stable strategy with another stable strategy within it (2), and (3) a stable strategy with an unstable strategy within it.

**Shape (2)** A structure with *m* number of 1-dimensional holes where $b_0 = 1, b_1 = m$, and $b_{n>1} = 0$ is an evolutionary stable strategy with *m* many stable clusters of one or more other strategies within it (Fig. 4(2)). It therefore reflects evolutionary coexistence. This corresponds to situations where stable clusters of a particular strategy exist within an ocean of another strategy, which is not unusual in spatial games (Helbing et al., 2010; Chen et al., 2014; Szolnoki and Perc, 2015; Hauert and Doebeli, 2004). A strategy structure with a single stable strategy within it is homotopically equivalent to a ring or annulus, as they both have one 1-dimensional hole. If we take the data points of the other stable strategies that occupy the holes, they are solid structures as defined in (1).

**Shape (3)** A structure with *m* number of 2-dimensional voids where $b_0 = 1, b_1 = 0, b_2 = m$, and $b_{n>2} = 0$ is an evolutionary stable strategy with *m* many patches of other unstable strategies within it (Fig. 4(3)). For instance, taking the data of strategy *S*, the proposal is that strategies that manage to invade the space of *S* for some time, but fail to remain stable in the long run, take the shape of a structure with 2-dimensional holes. We note that a strategy structure with a single unstable strategy within it is homotopically equivalent to a hollow sphere, as they both have one 2-dimensional void. Taking the data of the other strategies within the structure, they are solid structures as defined in (1), but since they are not stable we need to add a few assumptions.

The rigor of the three shapes is based on a three critical assumptions:

(1) Total occupancy – The first assumption is that the structures only hold if we assume that every space at every iteration is occupied by a strategy. Allowing empty spaces would drastically change the shape of the game, and even if the persistence of patches and clusters of strategies would be tractable in theory, the analysis would be susceptible to grave perturbations.

(2) Edge exclusion – The second is that in the case of shape (2) and (3), we assume that the strategies within the stable strategy used as data do not occupy edges or corners of the lattice. The reason is that a stable strategy within another stable strategy will not be detected if it is in the edge or corner of the lattice, as it will not constitute a 1-dimensional hole in the data. Similarly, an unstable strategy occupying the edges will not be detected as it fails to create a 2-dimensional void. A rationale for this rather strong assumption can be made with an appeal to the nature of edges. As noted in other spatial games (Masuda and Aihara, 2003; Lindgren and Nordahl, 1994), depending on the rules of interaction, edges and corners can offer a certain protection as players are more likely to avoid engagement with other competing strategies. For instance, in a game where players can interact with any of its 8 closest neighbors, players occupying edges will only be able to interact with 5, and the ones in corners only 3, as the space beyond the edge of the lattice is empty. This constitutes an unfair competitive advantage, or in some cases, disadvantage. Since it is unclear whether strategies in edges and corners would be stable or unstable at any other space in the lattice, they can be excluded with good reason.

(3) Iteration intervals of interest – The third assumption is that, in a PH analysis of a spatial game, we are only interested in the relevant iteration intervals of a certain strategy, meaning that they have a significant first appearance and end. The assumption arises as a response to the difficulty of telling the difference between a stable structure and an unstable one, particularly in respect to random initial conditions and games where there are no stable final states.

As observed in most spatial games, initial conditions such as the distribution of strategies and their location plays a crucial role for the outcome of a game, and different conditions can lead to significantly distinct outcomes (Perc et al., 2017; Nowak and May, 1992). To partly circumvent this issue, random initial conditions are often used to ensure that each strategy occupies the same amount of space. However, this alone does not guarantee that the strategies have an equal chance of survival, especially if there are three or more competing strategies. Perc et al. (2017) argues that random initial conditions, combined with a sufficiently large system size, still offers the best alternative as it provides a chance for each subsystem to emerge locally somewhere in the population, and later on gives the most stable subsystems a chance to invade the population as a whole. This presents a challenge for persistent homology, as random initial conditions give raise to immensely complicated yet trivial shapes. The pragmatic solution is to only analyze the significant subsystems that appear after the initial turmoil. After all, while it remains an open question where to draw the line between initial chaos and significant features, PH essentially provides a way to specify what qualifies as significance.

The assumption offers a similar pragmatic solution to the problem of absent final states. This issue does not target games that reliably reaches a stable and fixed state; whether finite or infinite, they sooner or later converge to the Nash equilibria. In such situations we can simply choose the first iteration of the frozen state as the final points of the data set used for PH analysis. However, for other games, including cyclical games (e.g., RPS games and disease models) and open-ended games (e.g., through the use of stochastic rules and mutation), we are left with no choice but to choose some reasonable end of the iteration interval that ideally reflects the success of the strategy. Interesting features in cyclical games would for instance be recurring shapes, and for more open-ended scenarios, the shape of a strategy existing over a relatively large number of iterations.

The third assumption also offers a way to deal with two particular problems that stems from shape (2) and (3). In the case of the former, a strategy other than the one used as data can be stable without creating a 1-dimensional hole in the data. As an example, consider a situation where, after an arbitrary number of initial iterations, mutant strategy *A* invades strategy *B* and remains stable until the end of the game. To solve this problem, we assume that in order for a structure of type (2) to contain other stable strategies within it, including mutants, we only consider the iteration interval of interest, namely the ones between the first significant appearance of the other strategy and the relevant ending state of the game. In this way, invading strategies that remain stable conform to the holes of shape (2).

In the case of the latter, we noted that in the case of shape (3), if we analyze the data of the strategies within the stable strategy, they will appear to be stable structures akin to shape (1), even though they are in fact unstable. However, in our definition of shape (1), we claimed that in order for a structure to be stable, it has to exist either at the entire course of the game or at the relevant interval of iterations. Similar to the case of shape (2), the relevant interval is once again between the first significant appearance of the strategy and some non-trivial final state. Consequently, features that disappear before the significant final state are therefore unstable.

Furthermore, the emergence of Shape (3) in simulation data implies that a strategy is not only evolutionary stable, but that it is also resistant to transient invasions. This observation is particularly helpful when we consider finite populations where there is a non-zero chance that a mutant strategy successfully invades and remain stable even if the strategy is opposed by natural selection in infinite populations (Neill, 2004; Nowak et al., 2004).

## 3. Evolutionary game models

In this section, we create two game models. The first is based on the standard Prisoners Dilemma with a few additional features. Importantly, it draws on insights from spatial games where stable clusters of a certain strategy can emerge and sustain in a sea of another dominant strategy (Helbing et al., 2010; Chen et al., 2014; Szolnoki and Perc, 2015; Hauert and Doebeli, 2004). Several extensive studies based on Prisoners Dilemma and Public Goods games have confirmed that population structure is beneficial for the persistence of cooperation, as it allows cooperators to form clusters that are less likely to be exploited by defectors (Hauert, 2006; Hauert and Szabo, 2003; Killingback et al., 2006; Nowak and May, 1992; Nowak et al., 2004). Additional features of our model include mutation and death by age in order to enable open-endedness (Lindgren and Nordahl, 1994), stochastic interaction to avoid the trivial patterns of deterministic cellular automata (Nowak and May, 1993), and memory to allow for strategies such as tit-for-tat (Axelrod and Axelrod, 1984).

The second game, Earth-Fire-Grass is based on a spatial SIRS disease model, as extensively studied in epidemiology (Li et al., 2009; Durrett, 1999; van Ballegooijen and Boerlijst, 2004). Similar models have also been used in the context of forestation and wildfire dynamics (Antonovsky et al., 1989; Durrett and Ma, 2018). Since many disease models involves cyclical dynamics, it also shares similarities with Rock-Paper-Scissors, as explored in behavioral and animal ecology (May and Leonard, 1975; Reichenbach et al., 2007), and found in models of the parasitic plant *Rhinanthus minor* (Cameron et al., 2009) as well as the side-blotched lizard *Uta stansburiana* (Sinervo and Lively, 1996). One interesting aspect of cyclical games is that they are often better described by evolutionary dynamics rather than NE, especially in the absence of stable NE (Goeree and Holt, 1999; Cason et al., 2010; Hoffman et al., 2015). In respect to a persistent homology analysis, the distinct feature of cyclical games is that, while they do not conform to the shapes described in the previous section, they might still provide interesting spatial information.

### 3.1. Prisoner's Dilemma model

The Prisoners Dilemma is the quintessential game for studying the evolution of cooperation, following the influential work by Axelrod (Axelrod and Axelrod, 1984). In our spatial model, each space of a finite 2D square lattice is occupied by a player. For each iteration, the following two phases occur:

In the first phase, every player is paired with one randomly selected neighbor out of the eight closest ones. Players use one of two actions, defect or cooperate, and the outcome is determined by the canonical payoff matrix:

|  |  | Player 2 | |
|---|---|---|---|
|  |  | Cooperate | Defect |
| Player 1 | Cooperate | $(R, R)$ | $(S, T)$ |
|  | Defect | $(T, S)$ | $(P, P)$ |

where $T$, defecting against cooperator $> R$, cooperate against cooperator $> P$, defecting against defector $> S$, cooperate against defector. Each player has a value that keeps track of its current score, and at start, all players are given the same starting score ($SS$) and maximum score ($MS$) that limits the current score. After each game, both players update their scores according to the payoff values, and the phase ends when every player has played.

The second phase governs the death, rebirth, and mutation of players. Players can die in two ways, either by having a score less than or equal to 0, or of old age. The chance of dying of old age is given by $(\lambda/\zeta) - 1$, where $\lambda$ is player age and $\zeta$ is the rate of senescence. For instance, if $\zeta = 10$, a player with $\lambda = 15$ has a 50% chance of dying, and 90% chance if $\lambda = 19$. If a player dies, it is either reborn as a randomly selected neighbor or mutates into a randomly selected strategy. The first represents reproduction proportionate to the local distribution of strategies, as the player inherits the strategy and current score value of the randomly selected neighbor (using the same pairing process as in the first phase). Mutation means that a player has a chance, given by probability $\mu$, to be reborn as a randomly selected strategy (out of the four possible) with score set to $SS$.

Four strategies are considered:

- Defector (D) – Always defects.
- Cooperator (C) – Always cooperates.
- Tit-for-tat (TFT) – After initial cooperation, it replicates the action done by the previous opponent (Axelrod and Axelrod, 1984).
- Anti-tit-for-tat (ATFT) – After initial defection, it does the opposite action done by the previous opponent (Yi et al., 2017).

Mutation and death by old age are implemented to allow for open-ended processes (Lindgren and Nordahl, 1994). The former enables mutant strategies to spontaneously invade. The latter ensures that players who only interact with strategies that keep their fitness $> 0$ will eventually die, further reducing the chance for the game to end in a frozen state. The complementary senescence value ($\zeta$) is used to prevent multiple players from dying at the same time, and a one step-memory is included to allow for one-shot versions of TFT and ATFT. The reason to include TFT in the model is that, in comparison with the other strategies, TFT will supposedly be resilient to defectors by defecting themselves, and will thus serve to exhibit spatial behavior relevant to our topological analysis. In fact, since conditionally cooperating players (e.g., our modified version of TFT) are likely to receive mutual cooperation payoff ($R$) within clusters of cooperators, and defect against neighboring defectors (receiving payoff $P$), it is possible for them to not only resist invasion from, but also spatially invade a population of defectors (this intuition is further explored in Alexander and Skyrms, 1999). The reason to include ATFT is to simply increase the complexity and spatial heterogeneity of the game (Mathiesen et al., 2011; Perc et al., 2017).

As each player only interacts with only one neighbor at each iteration, our game is different from conventional spatial models and Public Goods games where players play with all 4 (its von Neumann neighborhood) or 8 of its neighbors, and the payoff for one individual is the sum of those interactions (Perc et al., 2017; Nowak and May, 1993; Lindgren and Nordahl, 1994). The reason for this design is that stochastic pairing preserves the meaningful spatial structure while avoiding trivial determinism and grouping effects, i.e., that the outcome does not entirely depend on the initial configuration of the game and is less influenced by the particular choice of neighborhood. Consequently, it circumvents the kaleidoscopic type of spatial patterns that stem from deterministic group-wise interaction (Nowak and May, 1993; Lindgren and Nordahl, 1994). The model is also different from what is conventionally called stochastic Prisoners Dilemma, where the choice of action is determined probabilistically (Nowak, 1990; Grim, 1996). Rather, our model is stochastic in the sense that each player has an equal probability to play with any of its closest neighbors, and similarly, an equal chance to be overtaken by any of them at death. In turn, this makes our one-shot versions of TFT and ATFT relatively unconventional, as memory strategies are usually implemented in iterated Prisoner's Dilemma where players play with the same opponent repeatedly.

More precisely, for player $P$ and its set of neighbors $N(P) = \{N_1, N_2, \ldots, N_n\}$, $P$ has a $\frac{1}{|N(P)|}$ probability to be paired with any member of $N(P)$. $|N(P)| = 8$ for players not occupying edges or corners, 5 for players on edges, and 3 for players in corners. The payoff for a cooperator surrounded by defectors has then the probability $\frac{8}{8}$ to be $S$, and the payoff for a defector surrounded by defectors has the same probability to be $R$.

The process can be modeled more generally as a finite homogeneous Markov chain (Agapie et al., 2014), where the chance for a strategy $A$ at one lattice to change into strategy $B$ is given by the probability transition:

$$P[A \rightarrow B] = (X \cup Y) \cdot (Z + R)$$

where:

- $X$ is the probability that $A$ is paired with a strategy $K$ that would result in a death for $A$ by score $\leqslant 0$. It is given by the number of $K$ in the neighborhood $N(A)$ divided by the total number of neighbors in $N(A)$.
- $\cup$ is the numerical probability of mutually inclusive events, following the addition rule of probability such that $X \cup Y = X + Y - (X \cdot Y)$.
- $Y$ is the probability that $A$ dies of old age, given by the age of $A$ divided by the rate of senescence, $\frac{\lambda(A)}{\zeta}$.
- $Z$ is the probability that $A$ mutates into $B$, given by $\mu \frac{1}{Q}$, where $Q$ is the number of different strategies players can mutate into.
- $R$ is the probability that $A$ is overtaken by $B$ provided that it does not mutate. It is given by the number of $B$ in $N(A)$ divided by $|N(A)|$, which is then multiplied by $1 - \mu$.

In simple terms, it is the probability that $A$ dies and mutates into $B$ or is overtaken by neighbor $B$ (a more detailed equation is provided in Appendix A). The whole lattice, moving from one state $i$ to another $j$, is consequently the combined product of all the probability transitions $P[A \rightarrow B]$ of the entire lattice of a population with size $S$, yielding the transition matrix:

$$\sum_{j=1}^{S} P_{i,j} = 1$$

### 3.2. Earth-Fire-Grass model

Earth-Fire-Grass is a cyclical SIRS (Susceptible-Infectious-Recovered-Susceptible) model played on a finite 2D square lattice. Similar models have been used to illuminate spatial disease dynamics (Li et al., 2009; Gan et al., 2011; Ai and Albashaireh, 2014). It has three simple rules:

- Fire (F) burns grass (G)
- When fire burns out, determined by maximum age (MA), it turns into earth (E).
- When earth is fertile, determined by MA, it turns into grass.

Hence, Fire represents the infectious class, Grass the susceptible class, and Earth the recovered class. The mechanics differs from the Prisoners Dilemma model in two important ways. Firstly, if the age of a F individual is MA, it always changes into E, and similarly, if the age of an E reaches MA, it always changes into G. Secondly, fire only randomly selects among neighbors that are grass, excluding E and other F. This means that if there is a single grass $G_1$ in the neighborhood of some F, $G_1$ will turn into a F with probability $= 1$. This selection mechanism is the only source of randomness in the game.

## 4. Results

We did a number of persistent homology analyses on simulation data of the evolutionary game models and present six cases of special interest. For the Prisoners Dilemma (cases 1–5), the payoff values were fixed at $T = 2, R = 1, P = 0$, and $S = -1$, starting score at $SS = 2$, and maximum score at $MS = 4$. Fig. 5 shows baseline tests for seven different initial distributions of the Prisoners Dilemma without mutation or death by age.

In cases 3–5 we analyzed three particular iteration intervals from a single game played on a $7 \times 7$ lattice over 1000 iterations. Fig. 8 shows the distribution of strategies for every iteration of this game. Death by old age and mutation were only used for cases 3–5, where $\zeta = 20$ and $\mu = 0.05$.

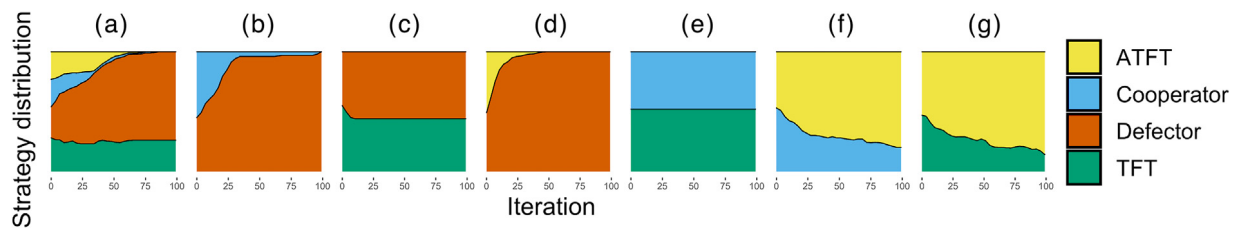

**Fig. 5.** The success of strategies in seven different initial distributions over 100 iterations on a 10x10 lattice, without mutation or death by age. (a) All four strategies, (b) defectors vs cooperators, (c) defectors vs TFT, (d) defectors vs ATFT, (e) cooperators vs TFT, (f) cooperators vs ATFT, and (g) TFT vs ATFT. As expected, Defect (D) is strategically dominant against every strategy except TFT. In cooperators vs ATFT (f) and ATFT vs TFT (g), ATFT eventually becomes the only surviving strategy. Note that although one might expect TFT and ATFT to be equally successful when paired against each-other (as they alternate between cooperation and defection), the one-shot nature of the stochastically decided interactions seems to favor ATFT (g).

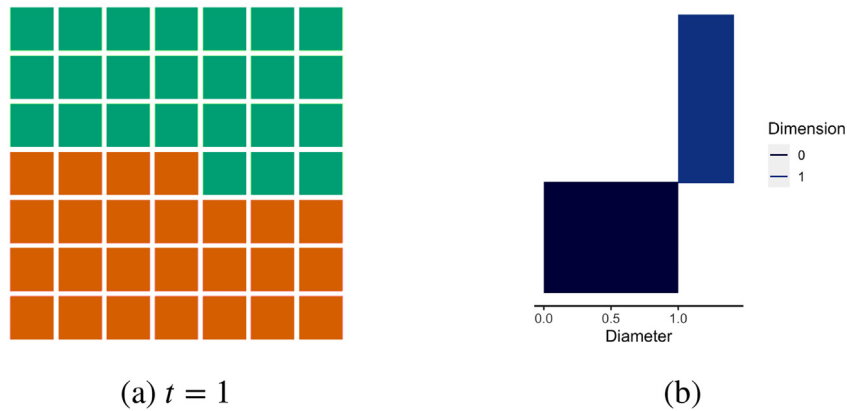

**Fig. 6.** (a) Initial distribution of the game in Case (1), where defect (red) and TFT (green) are each given half of the lattice. The persistent homology analysis (b) shows no significant features above $\epsilon = 1.41$.

### 4.1. Stable strategy without inner invasion

**Case (1)** We let two strategies, defect and TFT, play on a $7 \times 7$ lattice over 25 iterations. At start, the strategies were given half of the lattice each, divided vertically (Fig. 6(a)). After initial cooperation, TFT starts to defect against the defectors, effectively resulting in a frozen state. The PH analysis shows no significant features above distance $\epsilon = 1.41$ (Fig. 6(b)), meaning that it conforms to Shape (1) as defined in Section 2.2, where $b_0 = 1$ and $b_{n>0} = 0$. Using the data from either strategy yields the same results.

### 4.2. Stable strategy with stable clusters

**Case (2)** We let defect and TFT play on another prearranged $7 \times 7$ lattice over 30 iterations. Two clusters of TFT strategies were placed within a lattice occupied by defectors (Fig. 7(a)). The outcome mirrors the results from Case 1, with two strategies remaining in a frozen state over the entire course of the game (Fig. 7(b)-(c)). The PH analysis shows that two 1-dimensional holes persist between $\epsilon = 1.41$ and $2.23$ (the two blue lines in Fig. 7(d)). The single hole that persists above $\epsilon = 2.23$ reflects the merging of the two holes, corresponding to the euclidean distance between $P_1(0,0)$ and $P_2(1,2)$. Using the data of defectors, the structure conforms to Shape (2), with two 1-dimensional holes such that $b_0 = 1$, $b_1 = 2$, and $b_{n>1} = 0$, meaning that we have a stable strategy with two stable clusters of another strategy within it.

### 4.3. Stable strategy with unstable invasions

During the course of the 1000 iteration long game (Fig. 8), we found three intervals of particular interest, presented in Cases (3), (4), and (5).

**Case (3)** The third occurs between iteration $t = 596 - 621$ (Fig. 9(a)), where we find an example of a longer unsuccessful invasion. Using the data of the dominant defectors, the PH analysis accurately detects the event; the structure has one distinct 2-dimensional hole as indicated by the red line between $\epsilon = 1.41$ and $2.23$. This means that it conforms to Shape (3) in the sense that $b_0 = 1, b_1 = 0, b_2 = 1$, and $b_{n>2} = 0$.

**Case (4)** The fourth happens between iteration $t = 501 - 510$ (Fig. 9(b)), where we find two shorter unsuccessful invasions. Even if the invasions only last for two iterations each, the structure still conforms to Shape (3) but with two distinct 2-dimensional holes $b_2 = 2$, which affirms the robustness of the shapes.

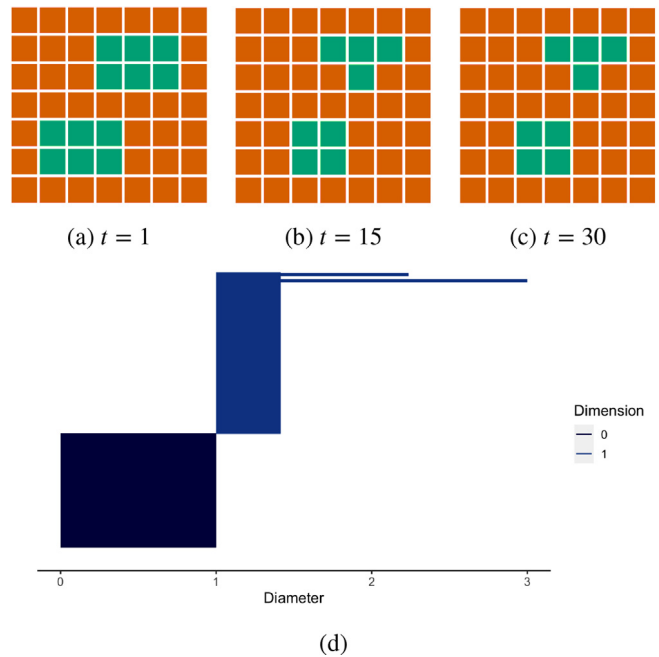

**Fig. 7.** (a), (b), and (c) show the lattice of Case (2) at the beginning (first iteration), middle (15th iteration), and end (30th iteration). (d) The PH analysis reveals two significant holes that persist between $\epsilon = 1.41$ and $2.23$, represented by the blue lines. The single hole persisting beyond $\epsilon = 2.23$ corresponds to the merging of the two holes into one. In order to more clearly show the number of significant features, elements in the PH analysis between $\epsilon = 0$ and $1.41$ are vertically shrunken by 50% and the ones above $1.41$ are vertically enlarged by 100%.

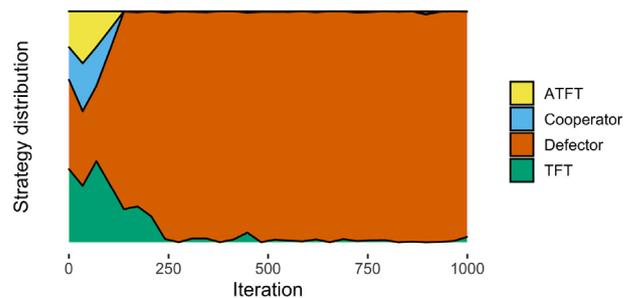

**Fig. 8.** The distribution of strategies over 1000 iterations as used in Case (3), (4), and (5). $SS = 2, MS = 4, \mu = 0.05, \zeta = 20$, with payoff values $T = 2 > R = 1 > P = 0 > S = -1$.

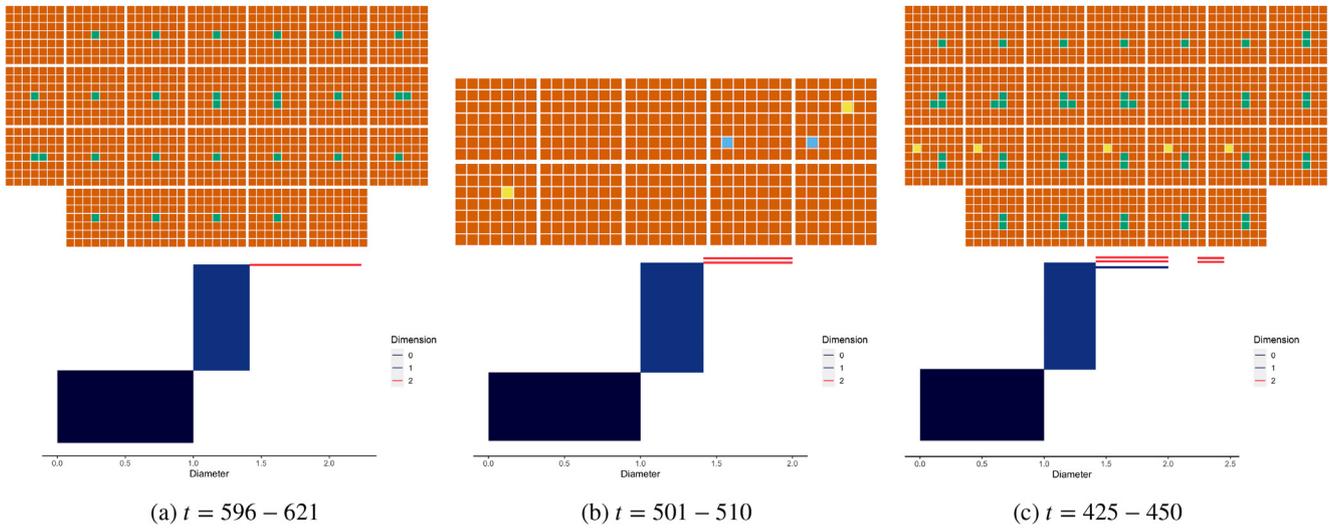

(a) $t = 596 - 621$       (b) $t = 501 - 510$       (c) $t = 425 - 450$

**Fig. 9.** Iteration prints and persistent homology analysis of Case 3 (a), 4 (b), and 5 (c). (a) A longer unsuccessful invasion corresponding to one distinct 2-dimensional hole ($b_2 = 1$) as described in Shape (3). (b) Two shorter unsuccessful invasions also conforming to Shape (3) but with two 2-dimensional holes ($b_2 = 2$). (c) One stable strategy (blue line) and two unsuccessful invasions (red lines), yielding a combination of Shape (2) and (3) in the sense that $b_1 = 1$ and $b_2 = 2$. In the PH figures, features between $\epsilon = 0$ and 1.41 are reduced vertically by 50% and enlarged by 100% above 1.41.

*4.4. Stable and unstable strategies*

**Case (5)** The fifth case occurs between iteration $t = 425 - 450$ (Fig. 9(c)). Within this range, we find examples of a stable group of TFT and two unstable invasion of ATFT within a lattice otherwise occupied by defectors. These events are adequately captured by the PH analysis; the two red lines between $\epsilon = 1.41$ and 2 reflects the two unsuccessful invasions, and the blue line between the same distance represents the stable cluster of TFT. In effect, the structure is a combination of Shape (2) and (3) in the sense that $b_0 = 1, b_1 = 1, b_2 = 2$, and $b_{n>2} = 0$. The red lines between $\epsilon = 2.23$ and 2.45 represents the creation of two insignificant combinations of the ATFT invasions above the merging point ($P_1(0,0,0)$ and $P_2(1,0,2)$).

*4.5. Cyclical stability*

**Case (6)** We ran two versions of the Earth-Fire-Grass game on a $12 \times 12$ lattice over 20 iterations. In both versions, four fire strategies were placed in the center of the lattice at start. The only difference between the two versions is that maximum age (MA) was set to 2 in the first (Fig. 10 in the second (Fig. 10(b)). As shown in the prints of the first game, the shorter MA allows fire to perpetually ignite neighboring grass in the center of the grid, leading to continuous and cyclical waves of strategies taking turn in dominating the population (Fig. 10(a)). By contrast, after a single wave of fire and earth, the higher MA quickly leads to a stable state of grass (Fig. 10 (b)). Using fire as data, the PH analyses illustrate how the shape differs between the two games; the first exhibiting a rich web of topological information (Fig. 11(a)), the second none at all (Fig. 11 (b)). Although cyclical games do not conform to the shapes described in Section 2.2, a PH analysis can determine whether any significant features arise at all (e.g. due to cyclical dynamics), or whether they do not.

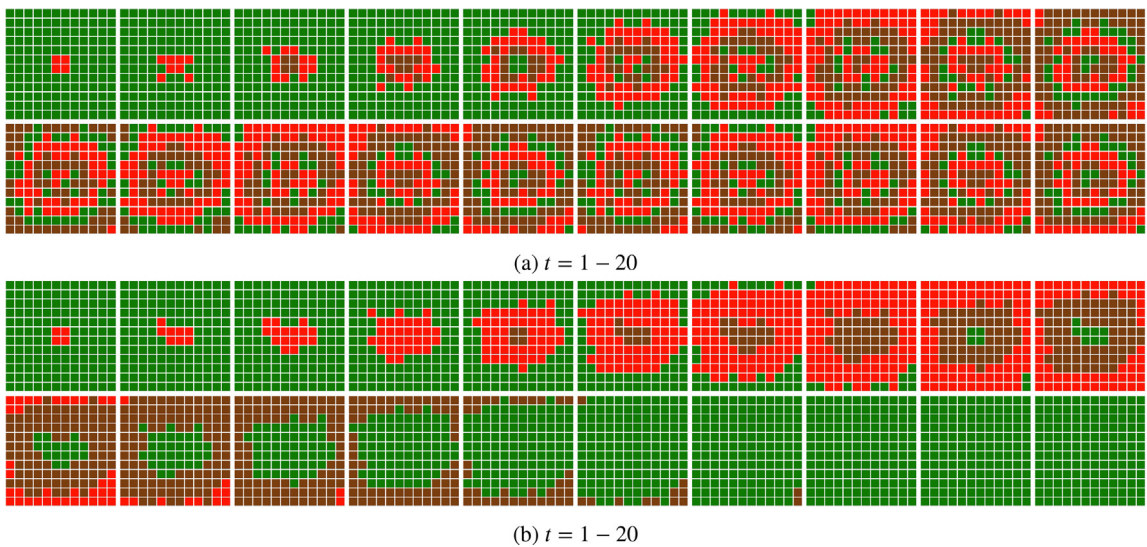

(a) $t = 1 - 20$

(b) $t = 1 - 20$

**Fig. 10.** Iteration prints of the two Earth-Fire-Grass scenarios used in Case (6), where $MA = 2$ for (a), and $MA = 4$ for (b). In the first version (a), fire (red) is able to continuously spread to neighboring grass (green) in the center of the lattice, which leads to long-lasting waves of spatial patterns spreading across the population. By contrast, in the second version (b), the game quickly ends up in a frozen state only populated by grass.

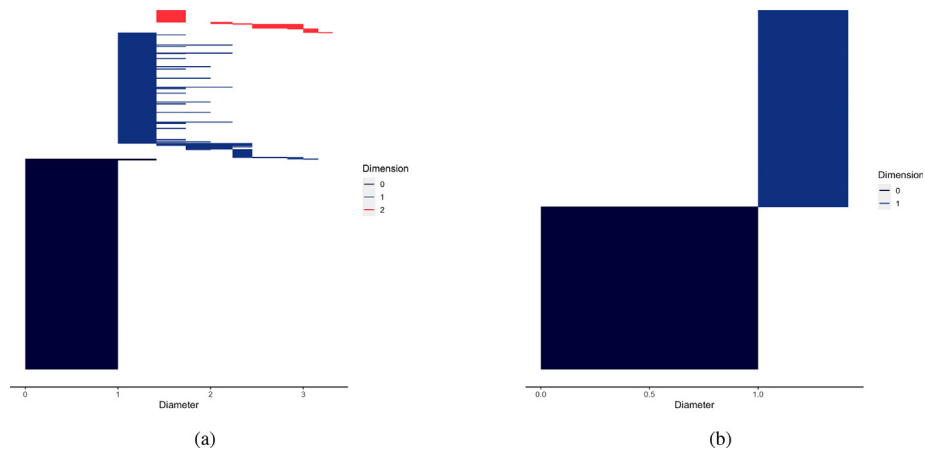

**Fig. 11.** PH analysis of the two versions of the Earth-Fire-Grass game used in Case (6). The first (a) displays a rich web of higher-dimensional features as a result of cyclical stability whereas the latter (b) shows no particular features of interest at all.

## 5. Discussion

We have introduced persistent homology as a novel methodology to the study of spatially extended games. While previous work have used PH to analyze time series of data in other contexts (Bonilla et al., 2020; Pereira and de Mello, 2015; Khasawneh and Munch, 2014), we have demonstrated how it is relevant for understanding spatial notions of evolutionary stability. In Section 2.2, we described three basic shapes that correspond to the stability and instability of strategies in a 2D lattice game (Fig. 4), and the experimental results show how a PH analysis accurately detects these shapes in a number of situations. For instance, it can tell us whether a strategy is stable (Section 4.1), has stable clusters within it (4.2), encompass unsuccessful invasions (4.3), or has combinations of these (4.4). Furthermore, we also demonstrated how PH can be used to track significant spatial features in cyclical games (4.5).

The major benefit with PH is that it provides a mathematically rigorous framework that can be used to both define and compute higher-order features of data in a manner that is independent of human observation, invariant to arbitrary parameter choices, and robust to noise. Our work demonstrates its relevance for spatial games in the sense that features that persist over spatial scales reflect the stability of strategies in a 2D lattice game. Regardless of the simplicity or complexity of a game, strategies succeed or fail to conform to shapes of a certain topological quality. Unlike other methods that study dynamics of spatial games, PH can tell us something meaningful about spatial dynamics while remaining neutral about the underlying structure and mechanisms.

A number of challenges remain for the application of PH to spatial games. As discussed in Section 2.2, one difficulty is to decide the appropriate iteration interval of interest used for analysis. This is particularly problematic for simulations using random initial distribution of strategies and games without stable final states, as these can potentially obscure the line between intractable chaos and significance. A related and more general problem is to deal with spatial chaos as such; although the robustness to noise is one of the key benefits of PH, highly intricate spatial patterns and configurations can have meaningful properties that are not easily tractable through the PH lens. To deal with the problem of random initial distribution, we proposed that one should analyze the relevant subsystems that appear after initial disorder and remain until the relevant end state of the game. We offered a similar pragmatic solution to the issue of absent final states, suggesting that one chooses iteration intervals that best reflect the nature of the game. As demonstrated in Case (5), relevant iteration intervals for open-ended games would ideally cover temporal stability; the shape of strategies that remain stable over a relatively large number of iterations. For cyclical games, relevant intervals would instead capture the recurring features that result from continuous turn-taking of strategic dominance, as shown in Case (6), or in other models that exhibits traveling waves. More importantly, while the precise details of these issues remain unsolved, PH provides a promising framework to address the open question of what ought to be regarded as spatial significance, e.g., by examining under what spatial scales certain features persist and form boundaries to others.

The emerging field of Topological Data Analysis is steadily providing new perspectives to an increasingly wide range of disciplines, and its application to spatial games presents several promising venues for future work. As different rules, parameter values, and population structures yield different outcomes, PH can be used to both conceptually define and computationally detect features of interest and their corresponding shapes in simulation data far beyond the ones described in this paper. In combination with other approaches, we therefore believe PH can provide novel insights to a diverse set of existing spatial games that study strategic interaction between individuals in socal science and biology, including Public Goods games (Szolnoki and Perc, 2010), Snowdrift games (Hauert and Doebeli, 2004), and War of Attrition (Bishop et al., 1978). The analysis of the Earth-Grass-Fire game demonstrates that PH can also shed light on cyclical dynamics found in spatial models of Rock-Paper-Scissors (Hoffman et al., 2015), disease spread (Ai and Albashaireh, 2014), forestation (Durrett and Ma, 2018), and wildfires (Antonovsky et al., 1989). To deal with the issue of edges and corners, games could be played on spheres or graphs where there is no strategic advantage or disadvantage of occupying edges and corners. Although we have only focused on one population structure the 2D lattice we believe there are other significant shapes to be found in other spatial systems. To that end, PH can be used in combination with evolutionary graph theory to illuminate how topological shapes arise in different population structures. Potential applications of particular promise are situations where emerging spatial properties such as patterns, waves, clusters, and boundaries play a key role in the success of certain strategies. Notable examples include games with cooperative clusters (Hauert and Szabo, 2003; Hauert, 2006; Killingback et al., 2006; Nowak and May, 1992), in-group/out-group effects (Bos et al., 2004), and games where players utilize the space in more sophisticated ways (Meloni et al., 2009).

Pattern formation has been extensively studied using reaction–diffusion (RD) models (Turing, 1990; Kondo and Miura, 2010) and positional information models (Wolpert, 1969; Green and Sharpe, 2015). In spatial games, RD models have been particularly useful in describing the underlying mechanisms that drive pattern formation and spatial heterogeneity (Wakano et al., 2009; Helbing, 2009). Furthermore, ordinary differential equations and RD models have been used to predict fixation probabilities and stationary distributions of stochastic models (Durrett and Levin, 1994; Durrett, 1999; Hwang et al., 2013). For instance, under the condition that players interact with a considerable part of the population, deterministic equations have been used to identify pattern formation and traveling front solutions in stochastic systems (Hwang et al., 2013). Similarly, it has been shown that the behavior of stochastic models can be determined from density equations by assuming that adjacent sites are independent (Durrett and Levin, 1994; Durrett, 1999). A related phenomenon is the concept of fundamental clusters, i.e., the smallest cluster sizes that can affect the long-term statistical behavior of the system (Killingback et al., 1999; Hauert, 2001). Hauert has described how fundamental clusters can be inferred statistically from growth criteria, and used to determine if mutant strategies are able to invade and survive in a population of other strategies (Hauert, 2001). However, while these approaches allow us to predict spatial dynamics and pattern formation in stochastic systems given specific conditions, assumptions, and information about the game, they are less useful in situations where the conditions do not hold, or when we do not know anything about the mechanisms of the underlying system. PH, on the other hand, can tell us whether significant features arise and sustain regardless of what we assume or know of the underlying system, provided that we have some data of the systems observed behavior and a homological concept of significance. We therefore believe there is a great theoretical and practical potential of combining and comparing PH with existing approaches to spatial dynamics and pattern formation. For instance, stochastic simulations or RD models can be used to produce spatial behavior and shapes which then can be analyzed through the topological lens, essentially creating a bridge between the generating mechanisms and the observed. In turn, this would allow us to illuminate more open problems regarding the role of patterns in complex and self-organizing systems (Landge et al., 2020). As a practical feature-detection tool, PH can automatically identify the conditions under which stable clusters arise and persist. To illustrate, one could continuously run a PH analysis in parallel with a simulation and change relevant parameter e.g., payoff values, update rules, population structure, reaction–diffusion forces until significant features appear. However, given the relatively high computational cost of PH, it remains unclear how it compares to other cluster analysis techniques and mere human observation. While a recent study shows that TDA outperforms traditional clustering algorithms in terms of accuracy (Almgren et al., 2017), further benchmarks could serve to show the advantage of PH in comparison to other point data processing methods and clustering algorithms such as OPTICS (Kriegel et al., 2011).

Furthermore, while a lot of work has highlighted the importance of space in the context of spatial games, it is also clear that space can influence the evolutionary dynamics in a variety of ways (Durrett and Levin, 1994). Therefore, besides combining PH with other methods, it is also possible to use it as a meta-methodology to systematically compare the results of different approaches to spatial modeling. For instance, recent work has demonstrated a model equivalence between the structure of a RD model and a positional information model (Vittadello and Stumpf, 2020), in effect providing a link between two distinct paradigms (Green and Sharpe, 2015).

To conclude, we have only scratched the surface of what TDA and PH can offer to our understanding of spatial games in particular and patterns in general. There is much potential for the method to provide novel insights for a wide variety of spatially extended systems in social science, physics, and biology. As Nowak and May wrote in 1993 after the arrival of spatial evolutionary games: "There is a new world to be explored" (Nowak and May, 1993).

**CRediT authorship contribution statement**

**Jakob Stenseke:** Conceptualization, Methodology, Formal analysis, Software, Visualization, Data curation, Writing - original draft, Writing - review & editing.

**Declaration of Competing Interest**

The authors declare that they have no known competing financial interests or personal relationships that could have appeared to influence the work reported in this paper.

**Acknowledgements**

This work was partially supported by the Wallenberg AI, Autonomous Systems and Software Program Humanities and Society (WASP-HS) funded by the Marianne and Marcus Wallenberg Foundation and the Marcus and Amalia Wallenberg Foundation. The author is grateful to his colleagues at the Department of Philosophy and the Department Cognitive Science at Lund University for insightful discussions and feedback on previous versions of the paper. The author would also like to thank the reviewers and editors for providing helpful comments that greatly improved the manuscript.

**Appendix A**

*Probability transition in the Prisoners Dilemma game*

$$P[A \to B] = \left( \frac{|K|}{|N(A)|} \cup \frac{\lambda(A)}{\zeta} \right) \left( \mu \frac{1}{Q} + (1 - \mu) \frac{|B|}{|N(A)|} \right)$$

where $|K|$ is the number of strategies $K$ in the neighborhood $N(A)$ and $|B|$ is the number of strategies $B$ in $N(A)$.

J. Stenseke

*Journal of Theoretical Biology 531 (2021) 110903*